\documentclass[amsmath, aps,prc,twocolumn,groupedaddress, showkeys, showpacs]{revtex4}
\usepackage{graphicx,epsfig, longtable}

\begin{document}

\title{Effective contact pairing forces from realistic calculations in infinite homogeneous nuclear matter}

\author{N.~Chamel}
\affiliation{Institut d'Astronomie et d'Astrophysique, CP-226, Universit\'e Libre de Bruxelles, 
1050 Brussels, Belgium}
\date{\today}

\begin{abstract}
Non-empirical effective contact pairing forces to be used in self-consistent 
mean-field calculations are presented. These pairing forces, constructed so as to 
reproduce exactly any given microscopic pairing gaps in infinite homogeneous nuclear matter 
for any isospin asymmetry, are given in analytical form. As a by-product, this work 
provides an analytical solution of the BCS gap equations which could be applied to 
describe various many-body systems. 
\end{abstract}


\maketitle

\section{Introduction}

Self-consistent mean-field calculations using effective interactions 
have been very successful in describing the properties and the 
dynamics of a wide range of nuclei~\cite{bhr03}. One of the most popular
effective interactions are zero-range interactions of the Skyrme type because of 
the fast numerical computations which can thus be performed~\cite{sto07}.
Even though Negele
and Vautherin~\cite{neg72} showed a long time ago how to obtain effective 
interactions using nuclear many-body methods, a more empirical approach has been 
usually followed. A specific parametric form of the effective interaction is postulated 
and the unknown parameters are determined \textit{a posteriori} so as to reproduce a set of 
nuclear data selected according to a specific purpose. The non-uniqueness of the fitting 
procedure has thus lead to a large number of different parametrizations. Some of them 
may yield very different predictions when applied outside the domain where they were 
fitted~\cite{sto03}. 

This situation is particularly unsatisfactory for nuclear astrophysical applications which 
require the knowledge of nuclear masses for nuclei so neutron rich that there is no hope of 
measuring them in the foreseeable future; such nuclei play a vital role in the r-process of 
nucleosynthesis~\cite{arn07} and are also found in the outer crust of neutron stars~\cite{pear09}. 
Extrapolations far beyond the neutron drip line are required for the description of the inner crust 
of neutron stars where nuclear clusters are embedded in a sea of superfluid neutrons~\cite{onsi08}. 
Such environments cannot be reproduced in the laboratory. However many astrophysical phenomena 
observed in neutron stars are precisely related to the physics of these outer layers~\cite{lrr}.
The need for more reliable extrapolations of these nuclear models has motivated recent efforts 
to construct non-empirical effective interactions and more generally microscopic nuclear energy 
density functionals~\cite{drut10}. Unfortunately such \textit{ab initio} nuclear energy density functionals 
able to reproduce existing experimental nuclear data with the same degree of accuracy as phenomenological 
interactions are not yet available. 

For the time being, a reasonable approach is to fit these phenomenological interactions with as 
many available nuclear data as possible, not only from experiments but also from realistic calculations
in infinite homogeneous nuclear matter. 
Very accurate nuclear mass models based on the 
Hartree-Fock-Bogoliubov (HFB) method with zero- and finite-range effective forces have thus been recently 
developed~\cite{cgp09,gor10}. 
In particular in our model HFB-17~\cite{gcp09,gcp09b}, we have achieved our best fit ever to essentially all the available 
experimental mass data, the rms deviation for the set of 2149 measured masses of nuclei with $N$ and $Z \ge$ 8~\cite{audi03} 
being only 0.581 MeV. Significant improvements compared to our previous models have been made possible by a better 
treatment of pairing correlations first introduced in Ref.~\cite{cha08} and extended in Ref.~\cite{gcp09,gcp09b}. The effective interaction 
used in the pairing channel was constrained to reproduce the $^1S_0$ pairing gaps as obtained by microscopic calculations 
using realistic nucleon-nucleon interactions. We have recently applied these effective forces to study the 
properties of the neutron superfluid phase in the inner crust of neutron stars~\cite{cha10}. 

The effective pairing interaction associated with our nuclear mass models requires the evaluation of one dimensional integrals. 
Although its numerical implementation is straightforward, using such interaction could become computationally costly for large scale
calculations such as the determination of fission barriers. In this paper, a more tractable 
expression of the pairing interaction is presented by calculating the underlying integrals analytically. 
We will consider effective interactions of the Skyrme type in the particle-hole channel. But results can 
be easily generalized to other kinds of interactions. 

\section{Effective density-dependent contact pairing force} 
\label{sec2}

The pairing interaction that we consider here acts only between nucleons of the same
charge state $q$ ($q=n$ or $p$ for neutron or proton respectively) and is given by
\begin{equation}
\label{eq.1}
v^{\rm pair}(\pmb{r_i}, \pmb{r_j})= 
\frac{1}{2}(1-P_\sigma)v^{\pi\, q}[\rho_n(\pmb{r}),\rho_p(\pmb{r})]~\delta(\pmb{r_{ij}})\, ,
\end{equation}
where $P_\sigma$ is the two-body spin-exchange operator, 
$\pmb{r_{ij}} = \pmb{r_i} - \pmb{r_j}$ and $\pmb{r} = (\pmb{r_i} + \pmb{r_j})/2$. 
Due to the zero range of this interaction, a cutoff has to be introduced in the 
gap equations in order to avoid divergences (for a review of the various prescriptions, 
see for instance Ref.~\cite{dug05}). In this work we include all single-particle 
states whose energy lies below $\lambda_q + \varepsilon_\Lambda$, where $\lambda_q$ is the 
chemical potential and $\varepsilon_\Lambda$ is an energy cutoff. Although microscopic calculations in 
semi-infinite nuclear matter suggest that pairing in nuclei is a surface 
phenomenon~\cite{bal04}, the density dependence of the pairing force is still poorly known. 
It is generally assumed that $v^{\pi\,q}[\rho_n,\rho_p]$ depends only on the 
isoscalar density $\rho=\rho_n+\rho_p$ and is often parametrized as~\cite{ber91}
\begin{equation}
\label{eq.2}
v^{\pi\,q}[\rho_n,\rho_p] = V_{\pi q}^\Lambda\left(1-\eta_q
\left(\frac{\rho}{\rho_0}\right)^{\alpha_q}\right) \, ,
\end{equation}
where $\rho_0$ is the nuclear saturation density while $V_{\pi q}^\Lambda$, $\eta_q$ and $\alpha_q$ are 
adjustable parameters. The superscript $\Lambda$ on $V_{\pi q}^\Lambda$ is to remind that the 
pairing strength depends very strongly on the cutoff used in the gap equations (in principle changing the cutoff 
modifies also the other parameters but the effects are usually found to be small). Effective interactions 
with $\eta_q=0$ ($\eta_q=1$) have been traditionally refered as volume (surface) pairing. 
The parameters in Eq.~(\ref{eq.2}) are usually fitted directly to experimental data. The standard
prescription is to adjust the value of the pairing strength $V_{\pi q}^\Lambda$ 
to the average gap in $^{120}$Sn~\cite{doba95}. However this does not allow 
an unambiguous determination of the remaining parameters $\eta_q$ and $\alpha_q$. 
Systematic studies of nuclei seem to favor a so-called mixed pairing 
with $\eta_q\sim 0.5$ and $1/2\lesssim\alpha_q\lesssim 1$~\cite{doba01,sam03}. 

Even though such forces have been widely applied in nuclear structure calculations with some 
success, they lack a direct connection with realistic nucleon-nucleon forces. The reliability 
of these forces far beyond the domain where they were fitted is therefore not guaranteed. In particular
pairing forces fitted only to finite nuclei generally yield unrealistic pairing 
gaps in infinite nuclear matter~\cite{tak94,cha08}, thus rendering their application to neutron-star crusts unreliable. 
Garrido et al.~\cite{gar99} proposed to determine the parameters of the pairing interaction in Eq.~(\ref{eq.2})
by fitting the $^1S_0$ pairing gaps in infinite symmetric nuclear matter as obtained by the realistic Paris 
potential in the BCS approximation. The parametric form Eq.~(\ref{eq.2}) has been recently extended by 
Margueron et al.~\cite{mar08} who introduced an isospin dependence in the pairing strength in order to 
reproduce the $^1S_0$ pairing gaps in both symmetric nuclear matter and pure neutron matter as obtained 
from Brueckner calculations~\cite{cao06}. This approach assumes that the pairing \emph{interaction} between two 
nucleons inside a nucleus is locally the same as the pairing interaction between two nucleons in 
infinite uniform matter. 
This should not be confused with the so-called ``local density approximation''
which supposes that the pairing \emph{field} $\Delta(\pmb{r})$ appearing in the HFB equations (see for instance
Appendix A of Ref.~\cite{cha08}) is locally the same in finite nuclei and in infinite nuclear matter. 
Even though the coupling to surface vibrations is expected to contribute to pairing~\cite{bri05}, a 
local pairing theory seems a reasonable first step~\cite{bul02,sat06}. Finite size effects can be subsequently 
taken into account by introducing density gradient terms in the expression of the effective pairing force 
as suggested in Ref.~\cite{bul04} in the general framework of the density functional theory. Effective 
pairing forces with density gradients have been implemented in Ref.~\cite{fay00}. 

The drawback of using a phenomenological pairing force in order 
to fit microscopic pairing gaps in infinite uniform matter is that the functional 
form of the associated pairing strength is not \textit{a priori} known. In particular different 
pairing gaps require different expressions of $v^{\pi\,q}[\rho_n,\rho_p]$ which can hardly 
be guessed due to the highly non-linear character of pairing correlations~\cite{dg04,zh10}.
It should also be remarked that the parameters of the pairing force depend on the 
effective interaction used in the particle-hole channel since effective masses
appear in the gap equations. Fitting the parameters of the effective interaction to essentially
all experimental nuclear masses both in the particle-hole and the particle-particle channels while 
simultaneously reproducing microscopic pairing gaps in infinite uniform matter would thus be an extremely onerous 
numerical task.

\section{Microscopically deduced contact pairing force} 
\label{sec3}

Actually as shown in Refs.~\cite{cha08,gcp09, gcp09b}, the density- and isospin dependence of the pairing strength 
can be calculated exactly for any given $^1S_0$ pairing-gap function $\Delta_q[\rho_n,\rho_p]$ 
by solving directly the HFB equations in infinite uniform matter yielding
\begin{equation}
\label{eq.3}
v^{\pi\,q}[\rho_n,\rho_p]=-\frac{8\pi^2}{I_q(\rho_n,\rho_p)}\left(\frac{\hbar^2}{2 M_q^*(\rho_n,\rho_p)}\right)^{3/2} 
\end{equation}
where $M_q^*(\rho_n,\rho_p)$ is the nucleon effective mass and 
\begin{equation}
\label{eq.4}
I_q(\rho_n,\rho_p)=\int_0^{\mu_q+\varepsilon_{\Lambda}}{\rm d}\xi 
\frac{\sqrt{\xi}}{\sqrt{(\xi-\mu_q)^2+\Delta_q(\rho_n,\rho_p)^2}}\, .
\end{equation}
The expression of $M_q^*(\rho_n,\rho_p)$ for Skyrme forces can be found for instance in Appendix A of 
Ref.~\cite{cha08}. 
In Eq.~(\ref{eq.4}), $\mu_q=\lambda_q-U_q$ is a reduced chemical potential 
($U_q$ being the mean field potential) which can be obtained by imposing 
the conservation of the nucleon particle number density
\begin{equation}
\label{eq.5}
\rho_q = \left(\frac{2 M_q^*}{\hbar^2}\right)^{3/2}\int_0^{+\infty}\frac{{\rm d}\xi}{4\pi^2}  \sqrt{\xi}\left(1-\frac{\xi-\mu_q}{\sqrt{(\xi-\mu_q)^2+\Delta_q^2}}\right) \, .
\end{equation}
However it is usually a very good approximation to replace $\mu_q$ in Eq.~(\ref{eq.4}) 
by the Fermi energy
\begin{equation}
\label{eq.6}
\varepsilon^{(q)}_{{\rm F}} = \frac{\hbar^2 k_{{\rm F}q}^2}{2 M_q^*} \, ,
\end{equation}
where $k_{{\rm F}q}= (3\pi^2 \rho_q)^{1/3}$ is the Fermi wave number. The approximation 
$\mu_q\simeq \varepsilon^{(q)}_{{\rm F}}$ holds provided $\Delta_q \ll \varepsilon^{(q)}_{{\rm F}}$ 
which is typically the case~\cite{lom99}. 

Using Eq.~(\ref{eq.3})
instead of a phenomenological expression such as Eq~(\ref{eq.2}) guarantees that in 
infinite uniform matter the microscopic pairing gaps $\Delta_q(\rho_n,\rho_p)$ will 
be exactly reproduced. It also ensures that the pairing strength is suitably renormalized
for any change in the energy cutoff. The price to be paid is the evaluation of the one dimensional
integral, Eq~(\ref{eq.4}), for the neutron density $\rho_n(\pmb{r})$ and proton density
$\rho_p(\pmb{r})$ at all points inside the nucleus. The computational cost becomes significant in 
fully self-consistent 2D and 3D calculations~\cite{pear09}. However we will now show that these 
numerical integrations can be avoided by making use of the weak-coupling approximation. 

The integrand in Eq.~(\ref{eq.4}) is all the more peaked around 
$\xi=\mu_q \simeq \varepsilon^{(q)}_{{\rm F}}$ as $\Delta_q$
is small compared to $\varepsilon^{(q)}_{{\rm F}}$. This suggests to expand 
the integrand in powers of $\Delta_q/\varepsilon^{(q)}_{{\rm F}}$ leading to
\begin{equation}
\label{eq.7}
I_q=\sqrt{\varepsilon^{(q)}_{\rm F}}\sum_{n=0}^{+\infty} \tilde{I_q}^{(n)}
\end{equation}
where the dimensionless coefficients $\tilde{I_q}^{(n)}$ are defined by 
\begin{equation}
\label{eq.8}
\tilde{I_q}^{(n)}=\frac{(2n)!}{(1-2n)(n!)^2}\left(\frac{-\Delta_q}{4\varepsilon^{(q)}_{\rm F}}\right)^n
\int_{-\varepsilon^{(q)}_{\rm F}/\Delta_q}^{\varepsilon_\Lambda/\Delta_q}\frac{x^n {\rm d}x }{\sqrt{1+x^2}}\, .
\end{equation}
In the weak-coupling approximation $\Delta_q \ll \varepsilon^{(q)}_{\rm F}$ and
$\Delta_q \ll \varepsilon_\Lambda$, 
only the first coefficient in Eq.~(\ref{eq.7}) is usually retained, i.e.  
$I_q\simeq \sqrt{\varepsilon^{(q)}_{\rm F}}\tilde{I_q}^{(0)}$. 
This approximation is equivalent in taking a constant density of single-particle states in 
the gap equations. Calculting the integral $\tilde{I_q}^{(0)}$ and keeping lowest order 
terms in $\Delta_q/\varepsilon^{(q)}_{\rm F}$ and $\Delta_q/\varepsilon_\Lambda$, 
Eq.~(\ref{eq.8}) thus yields
\begin{equation}
\label{eq.9}
\tilde{I_q}^{(0)}\simeq \log\left(\frac{4\varepsilon^{(q)}_{\rm F}\varepsilon_\Lambda}{\Delta_q^2}\right)\, .
\end{equation}
Inserting 
Eq.~(\ref{eq.9}) in Eq.~(\ref{eq.3}) and solving for $\Delta_q$ leads to the familiar expression (note 
that $v^{\pi\,q}<0$)
\begin{equation}
\Delta_q^{(0)}=2\sqrt{\varepsilon^{(q)}_{\rm F}\varepsilon_\Lambda}\exp\left(\frac{2\pi^2\hbar^2}{v^{\pi\,q} M_q^*k_{{\rm F}q}}\right)\, . 
\end{equation}
Even though the weak-coupling approximation provides good results in the case of conventional 
BCS superconductivity~\cite{bcs57}, it is less accurate in the nuclear context because of the large number of states involved 
in the pairing mechanism. Nevertheless the higher-order coefficients in Eq.~(\ref{eq.8}) can be easily evaluated. 
Calculating the integrals $\tilde{I_q}^{(n)}$, keeping as before lowest-order terms in $\Delta_q/\varepsilon^{(q)}_{\rm F}$ 
and $\Delta_q/\varepsilon_\Lambda$, and summing all coefficients, we find 
\begin{equation}
\label{eq.10}
\sum_{n=1}^{+\infty} \tilde{I_q}^{(n)}\simeq2\sqrt{1+y}-2\log\left[\frac{1}{4}\left(1+\sqrt{1+y}\right)\right]-4\, ,
\end{equation}
where $y=\varepsilon_\Lambda/\varepsilon^{(q)}_{\rm F}$. Adding Eqs.~(\ref{eq.9}) and (\ref{eq.10}) 
in Eq.~(\ref{eq.7}) leads to
\begin{equation}
\label{eq.11}
I_q=\sqrt{\varepsilon^{(q)}_{\rm F}}\biggl[ 2\log\left(\frac{2\varepsilon^{(q)}_{\rm F}}{\Delta_q}\right)+ \Lambda\left(\frac{\varepsilon_\Lambda}{\varepsilon^{(q)}_{\rm F}}\right) \biggr]
\end{equation}
with
\begin{equation}
\label{eq.12}
\Lambda(x)=\log (16 x) + 2\sqrt{1+x}-2\log\left(1+\sqrt{1+x}\right)-4\, .
\end{equation}
The highly-non linear character of the pairing phenomenon is evident in Eqs.~(\ref{eq.3}) and (\ref{eq.11}). These equations 
also show that the density dependence of the pairing strength is intimately related to the choice of the pairing cutoff. 
The contribution of the latter is entirely contained in the function $\Lambda(x)$. It can also be seen from Eqs.~(\ref{eq.3}) 
and (\ref{eq.11}) that the pairing strength vanishes whenever the pairing gap $\Delta_q$ goes to zero at \emph{finite} density, 
as expected. 

Note that as a by-product we have also obtained a more accurate expression of the pairing gap. 
Substituting Eq.~(\ref{eq.11}) in Eq.~(\ref{eq.3}) yields
\begin{equation}
\label{eq.13}
\Delta_q=\Delta_q^{(0)}\exp\left(\frac{1}{2}\Lambda(y)\right)y^{-1/2}\, ,
\end{equation}
where $y=\varepsilon_\Lambda/\varepsilon^{(q)}_{\rm F}$. 
\begin{figure}
\includegraphics[scale=0.3]{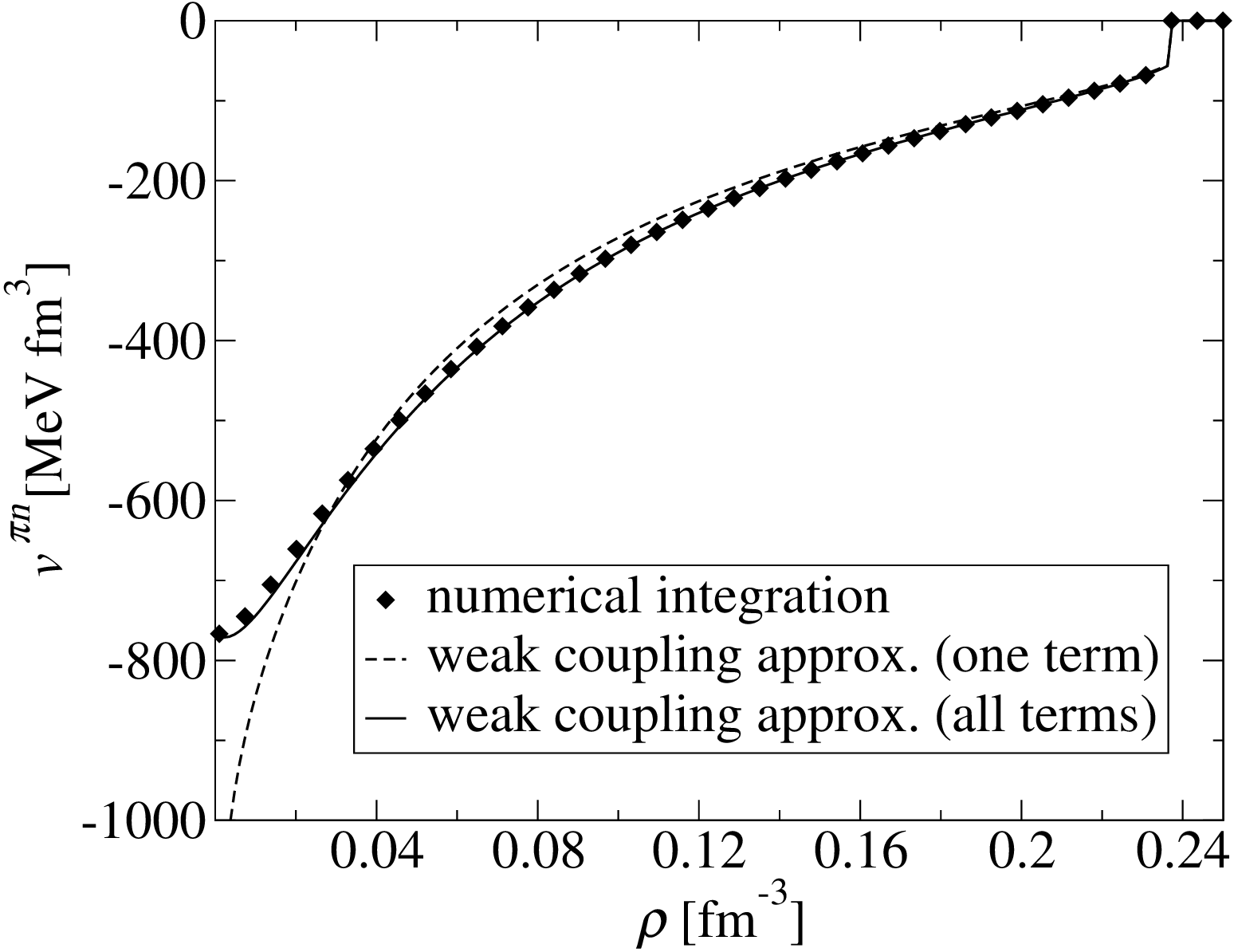}
\caption{Neutron pairing strength $v^{\pi\,n}[\rho_n,\rho_p]$, as defined by Eq.(\ref{eq.3}), 
vs nucleon number density $\rho$ in symmetric nuclear matter using the effective interaction underlying the nuclear 
mass model HFB-17~\cite{gcp09, gcp09b}. The symbols are the results of numerically integrating Eq.~(\ref{eq.4}).
The dashed and solid lines were obtained using either Eq.~(\ref{eq.9}) or Eq.~(\ref{eq.11}) respectively.}
\label{fig1}
\end{figure}

\begin{figure}
\includegraphics[scale=0.3]{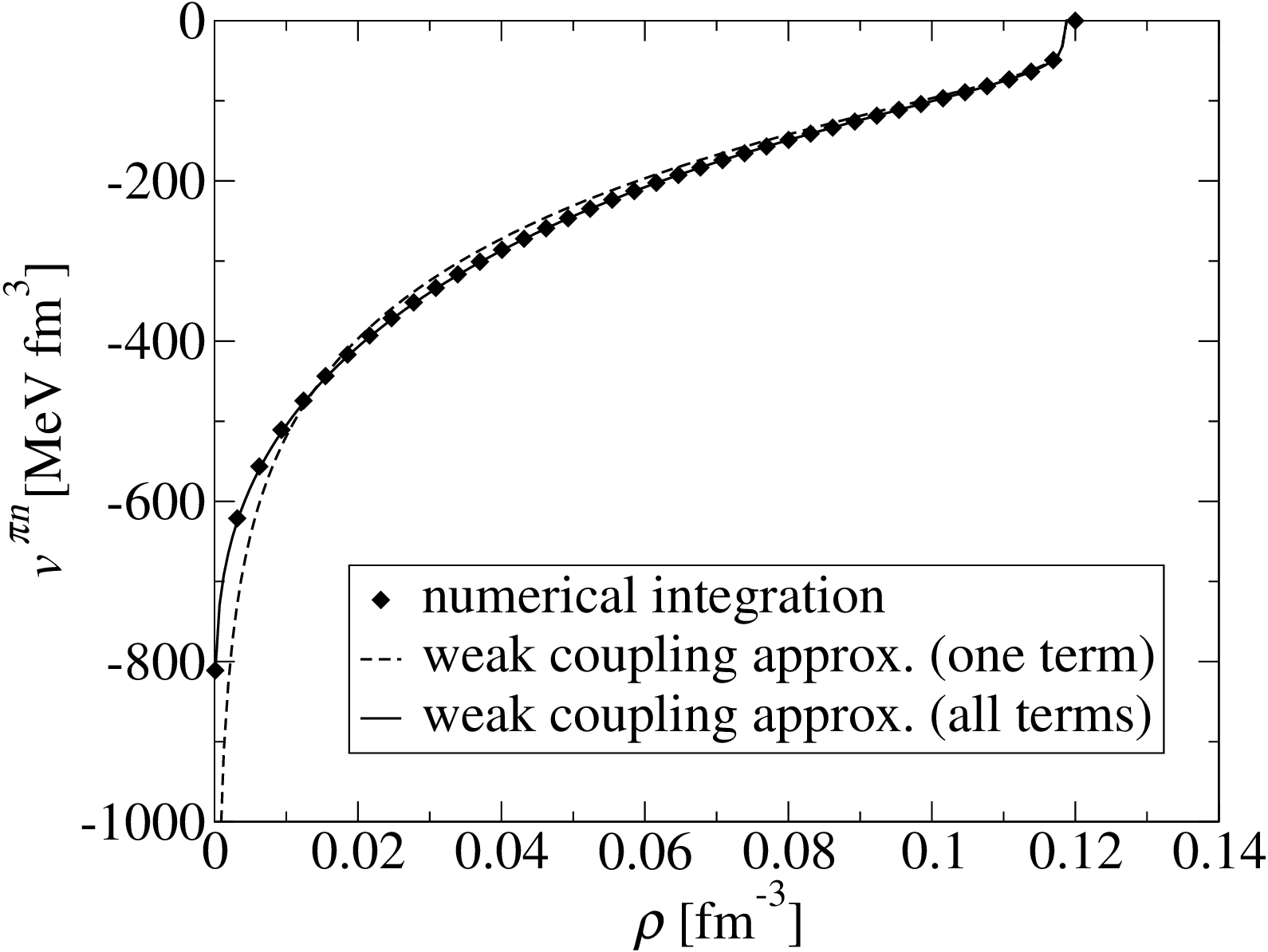}
\caption{Same as Figure~\ref{fig1} but for pure neutron matter.}
\label{fig2}
\end{figure}

We have tested the validity of the analytical expression given by Eq.~(\ref{eq.11}) as compared to the 
numerical integration of Eq.~(\ref{eq.4}) using the effective interactions underlying our HFB-17 
nuclear mass model~\cite{gcp09, gcp09b}. The pairing force used in this model was adjusted so as
to reproduce the $^1$S$_0$ pairing gaps in both symmetric nuclear matter and pure neutron matter
as obtained from Brueckner calculations using the Argonne $V18$ nucleon-nucleon potential~\cite{cao06}. 
The pairing cutoff was fixed at $\varepsilon_\Lambda=16$ MeV. As can be seen 
in Figures~\ref{fig1} and \ref{fig2}, Eqs.~(\ref{eq.3}) and (\ref{eq.11}) provide a very good estimate of 
the pairing strength for all densities and different isospin asymmetries. The figures also show that keeping 
only the first term in Eq.~(\ref{eq.7}) with $\tilde{I_q}^{(0)}$ given by Eq.~(\ref{eq.9}), leads to fairly good 
results except at low densities. 

\begin{figure}
\includegraphics[scale=0.3]{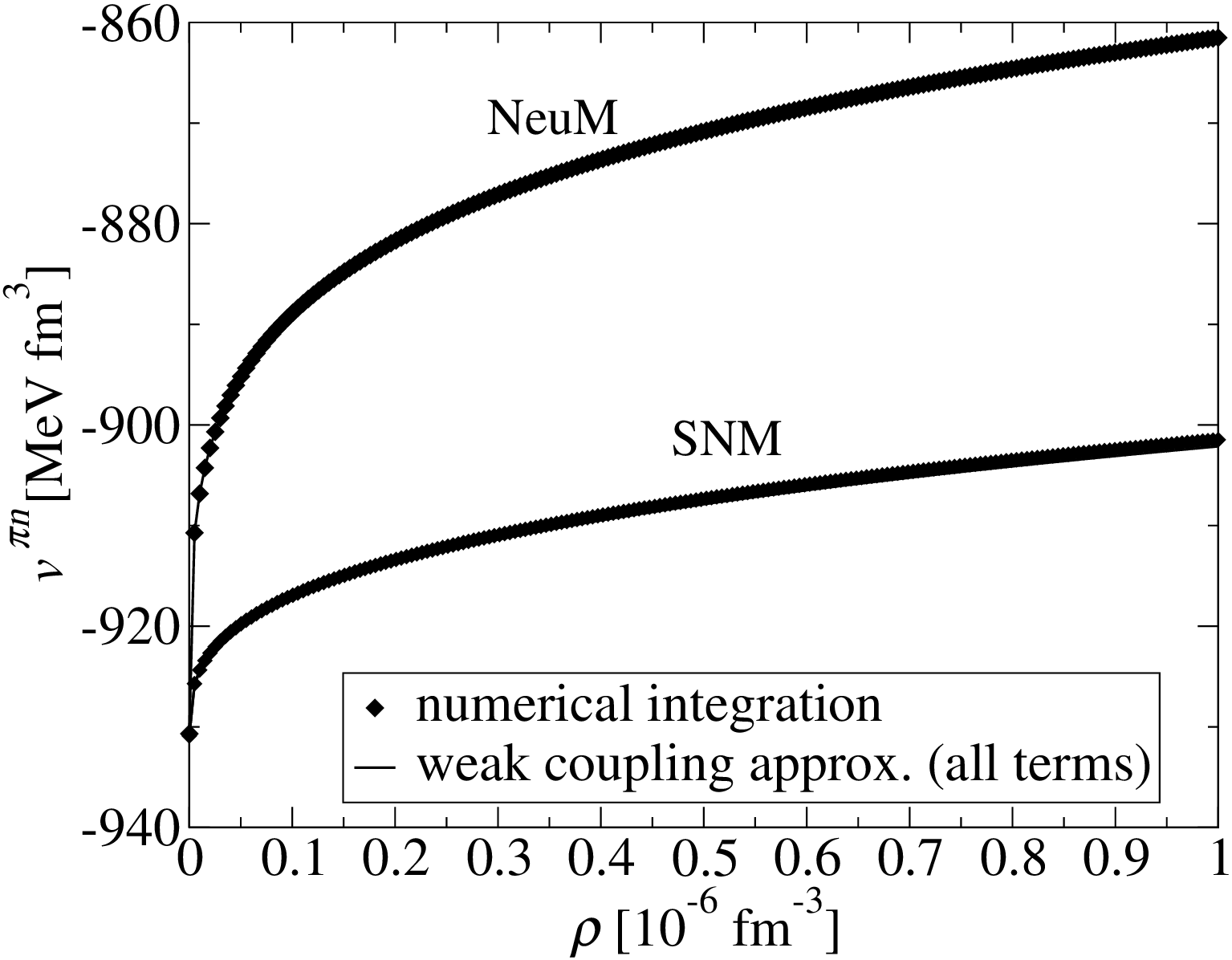}
\caption{Neutron pairing strength $v^{\pi\,n}[\rho_n,\rho_p]$, as defined by Eq.(\ref{eq.3}), 
vs nucleon number density $\rho$ in symmetric nuclear matter (SNM) and pure neutron matter 
(NeuM) using the effective interaction underlying the nuclear mass model HFB-17~\cite{gcp09, gcp09b}. 
The symbols are the results of numerically integrating Eq.~(\ref{eq.4}) while the solid line was obtained 
from Eq.~(\ref{eq.11}).}
\label{fig3}
\end{figure}

At very low densities $\rho\rightarrow0$, both the pairing gap $\Delta_q$ and the chemical potential vanish so that 
the integral $I_q$ is then simply given by $2\sqrt{\varepsilon_\Lambda}$. It is easily seen from Eq.~(\ref{eq.3}) that the pairing strength remains 
finite in this limit and is given by 
\begin{equation}
\label{eq.15}
v^{\pi\,q}[\rho\rightarrow0]=-\frac{4\pi^2}{\sqrt{\varepsilon_\Lambda}}\left(\frac{\hbar^2}{2 M_q}\right)^{3/2} \, ,
\end{equation}
assuming that $M_q^*(\rho\rightarrow0)=M_q$. 
The weak-coupling expression, given by Eqs.~(\ref{eq.3}) and (\ref{eq.11}), tends to the exact limit as shown in Figure~\ref{fig3}, 
whereas Eq.~(\ref{eq.9}) leads to a divergence (the corresponding pairing strength lies far below the range shown in 
Figure~\ref{fig3}). The reason for this discrepancy lies in the underlying assumption of a constant density of s.p. states, 
which is strongly violated at low energies $\varepsilon$ where the density of states decreases as $\sqrt{\varepsilon}$. 

\begin{figure}
\includegraphics[scale=0.3]{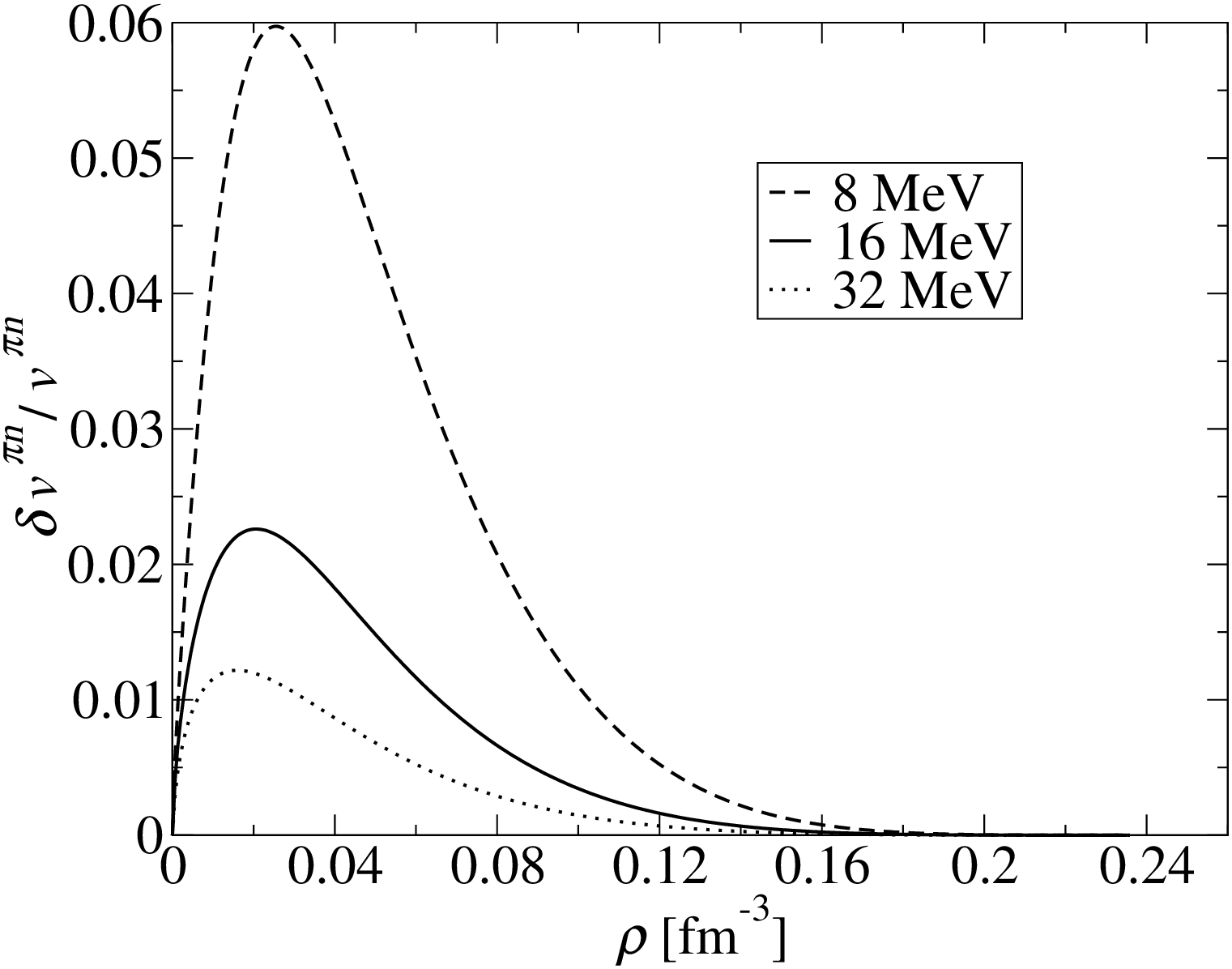}
\caption{Relative error between the neutron pairing strength $v^{\pi\,n}[\rho_n,\rho_p]$ 
calculated numerically and that obtained from the weak-coupling approximation, vs nucleon 
number density $\rho$. The different curves correspond to different values of the 
pairing cutoff $\varepsilon_\Lambda$ indicated in the plot. Calculations
were carried out in symmetric nuclear matter using the effective interaction underlying the 
nuclear mass model HFB-17~\cite{gcp09, gcp09b}.
}
\label{fig4}
\end{figure}

\begin{figure}
\includegraphics[scale=0.3]{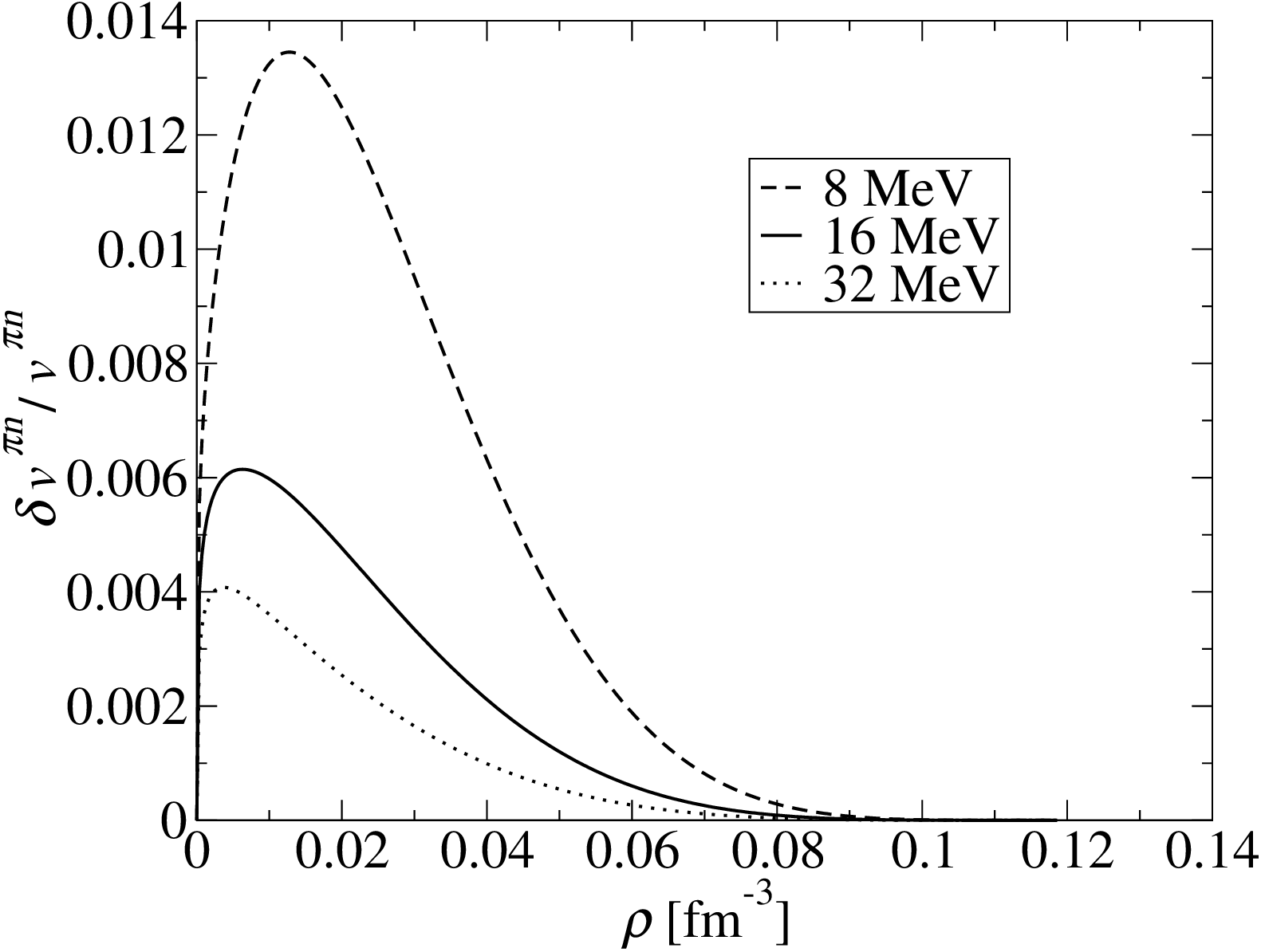}
\caption{Same as Fig.~\ref{fig4} but in neutron matter. 
}
\label{fig5}
\end{figure}

The value of the pairing cutoff $\varepsilon_\Lambda$ can be a priori arbitrarily chosen. However it has been argued~\cite{esb97,gar99} 
that in the limit $\rho\rightarrow0$, the pairing strength should coincide with the bare force in the $^1$S$_0$ channel, which 
in turn is determined by the experimental $^1$S$_0$ nucleon-nucleon phase shifts. According 
to Eq.~(\ref{eq.15}), specifying $v^{\pi\,q}[\rho\rightarrow0]$ is equivalent to specifying the pairing cutoff $\varepsilon_\Lambda$. 
As shown in Fig.~1 of Ref.~\cite{esb97}, the optimal value of the cutoff is $\varepsilon_\Lambda\sim 7-8$ MeV (note that 
$\varepsilon_\Lambda$ is half the cutoff used in Ref.~\cite{esb97}). 
Using the experimental phase shifts would thus remove the only free remaining parameter. On the other hand, 
choosing such a low cutoff can deteriorate the precision of the weak-coupling approximation, since Eq.~(\ref{eq.11}) was 
obtained assuming $\Delta_q\ll\varepsilon_\Lambda$. We have therefore calculated the relative error between the pairing strength 
calculated numerically and that obtained from Eqs.~(\ref{eq.3}) and (\ref{eq.11}), for different values of the pairing cutoff. 
Results are shown in Figs.~\ref{fig4} and \ref{fig5} for symmetric nuclear and pure neutron matter respectively. We have checked that the 
differences come solely from the weak-coupling approximation, and not from the numerical method used to solve Eq.~(\ref{eq.4}). 
It can be seen that for the 
effective interactions underlying our HFB-17 nuclear mass model~\cite{gcp09, gcp09b}, the weak-coupling approximation is quite accurate 
since the largest relative error is of a few percent only for $\varepsilon_\Lambda=8$ MeV. The precision is higher in neutron matter 
than in symmetric matter because the $^1$S$_0$ pairing gap of neutron matter is smaller than that of symmetric matter. Note that 
the pairing gaps we adopted in our HFB-17 mass model are rather large (the maximum of the gap function $\Delta_q(\rho_n,\rho_p)$ is about 
$4.8$ MeV in symmetric matter and $2.6$ MeV in neutron matter). These gaps were obtained from Brueckner calculations including 
medium polarization effects but without self-energy corrections~\cite{cao06}. When both effects are taken into account, the 
maximum pairing gaps are much lower~\cite{cao06}. The precision of the weak-coupling approximation would thus be even better 
with such gaps. In any case, if the constraint of reproducing the $^1$S$_0$ nucleon-nucleon phase shifts is released, one is 
free to adjust the cutoff so as to achieve a better precision, as illustrated in Figs~\ref{fig4} and \ref{fig5}.

In the more general context of the nuclear energy density functional theory, the HFB equations can be obtained 
by minimizing the energy with respect to the normal and pairing density matrices for a fixed average
number of neutrons and protons~\cite{doba96} (the link between different formulations of the HFB equations
is discussed for instance in Ref.~\cite{cha08}). All that is needed is therefore the specification of 
the nuclear energy density functional. The pairing energy density associated with an effective pairing force 
defined by Eq.~(\ref{eq.1}), is given by
\begin{equation}
\label{eq.16}
\mathcal{E}_{\rm pair}(\pmb{r})=\frac{1}{4} \sum_{q=n,p} v^{\pi\, q} [\rho_n(\pmb{r}),\rho_p(\pmb{r})] \tilde{\rho}_q(\pmb{r})^2 \, ,
\end{equation}
where $\tilde{\rho}_q(\pmb{r})$ is the so-called local pairing density. The pairing field appearing in the 
HFB equations is then defined by 
\begin{equation}
\label{eq.17}
\Delta_q(\pmb{r})\equiv\frac{\partial\mathcal{E}_{\rm pair}(\pmb{r})}{\partial\tilde{\rho}_q(\pmb{r})}\, .
\end{equation}
Because $\mathcal{E}_{\rm pair}(\pmb{r})$ depends on the nucleon densities, it contributes also to the mean-field
potentials (see for instance Appendix A of Ref.~\cite{cha08}). Local pairing functionals can be constructed from 
infinite nuclear-matter calculations using Eqs.~(\ref{eq.3}), (\ref{eq.11}) and (\ref{eq.12}). The construction
of non-empirical (and non-local) pairing functionals using realistic nucleon-nucleon interactions have 
been discussed in Ref.~\cite{les09} (see also Ref.\cite{drut10} for a review). 

\section{Conclusions}

In conclusion, we have shown how to construct non-empirical effective contact pairing forces (or equivalently 
local nuclear pairing energy density functionals) using any given $^1$S$_0$ pairing-gap functions $\Delta_q(\rho_n,\rho_p)$ 
obtained from microscopic calculations in infinite uniform nuclear matter with realistic nucleon-nucleon potentials. 
The resulting parameter-free pairing forces, embodied in Eqs.~(\ref{eq.3}), (\ref{eq.11}) and (\ref{eq.12}), 
can be easily implemented in nuclear-structure calculations. These forces could be helpful for understanding 
the origin of pairing in finite nuclei, and could be used to estimate the importance of bulk contribution as 
compared to finite-size effects. Alternatively, the analytical solution of the BCS gap equations which we have 
obtained in Eq.~(\ref{eq.13}), can be applied to calculate the pairing gaps in nuclear matter for any given contact 
pairing force (provided the conditions for the weak-coupling approximation remains valid). 
Even though we have been interested in nuclear pairing, the present results could be easily adapted to other contexts.

\begin{acknowledgments}
This work was financially supported by FNRS (Belgium), by the Communaut\'e fran\c{c}aise de Belgique 
(Actions de Recherche Concert\'ees) and by CompStar, a Research Networking Programme of the 
European Science Foundation. The author is grateful to S. Goriely and J.M. Pearson for valuable comments. 
\end{acknowledgments}

\end{document}